\documentclass{article}
\usepackage{eqalignno}

 2%
\font\sc=cmcsc10%
\font\tenmib=cmmib10 \font\eightmib=cmmib8
\font\sevenmib=cmmib7\font\fivemib=cmmib5
\font\ottoit=cmti8\font\fiveit=cmti5\font\sixit=cmti6
\font\fivei=cmmi5\font\sixi=cmmi6\font\ottoi=cmmi8
\font\eightrm=cmr8
\font\fiverm=cmr5\font\sixrm=cmr6%
\font\ottosy=cmsy8\font\sixsy=cmsy6\font\fivesy=cmsy5
\font\ottoex=cmex8\font\sixex=cmex6\font\fiveex=cmex5
\font\ottobf=cmbx8\font\sixbf=cmbx6\font\fivebf=cmbx5%
\font\ottocss=cmcsc8%
\def\ottopunti{%
\textfont0=\eightrm\scriptfont0=\sixrm\scriptscriptfont0=\fiverm%
\def\rm{\fam0 \eightrm}%
\textfont1=\ottoi\scriptfont1=\sixi\scriptscriptfont1=\fivei%
\def\mit{\fam1 \ottoi}%
\textfont2=\ottosy\scriptfont2=\sixsy\scriptscriptfont2=\fivesy%
\def\cal{\fam2 \ottosy}%
\textfont3=\ottoex\scriptfont3=\sixex\scriptscriptfont3=\fiveex%
\def\calx{\fam3 \ottoex}%
\textfont4=\ottocss\scriptfont4=\sc\scriptscriptfont4=\sc%
\def\csc{\fam4 \ottocss}%
\textfont5=\eightmib\scriptfont5=\sevenmib\scriptscriptfont5=\fivemib%
\textfont6=\ottoit\scriptfont6=\sixit\scriptscriptfont6=\fiveit%
\def\it{\fam6\ottoit}%
\textfont7=\ottobf\scriptfont7=\sixbf\scriptscriptfont7=\fivebf%
\def\bf{\fam7\ottobf}%
\baselineskip5mm \rm}%
\def\nota{\ottopunti\baselineskip10pt}%
\mathchardef\Ba   = "050B  
\mathchardef\Bb   = "050C  
\mathchardef\Bg   = "050D  
\mathchardef\Bd   = "050E  
\mathchardef\Be   = "0522  
\mathchardef\Bee  = "050F  
\mathchardef\Bz   = "0510  
\mathchardef\Bh   = "0511  
\mathchardef\Bthh = "0512  
\mathchardef\Bth  = "0523  
\mathchardef\Bi   = "0513  
\mathchardef\Bk   = "0514  
\mathchardef\Bl   = "0515  
\mathchardef\Bm   = "0516  
\mathchardef\Bn   = "0517  
\mathchardef\Bx   = "0518  
\mathchardef\Bom  = "0530  
\mathchardef\Bp   = "0519  
\mathchardef\Br   = "0525  
\mathchardef\Bro  = "051A  
\mathchardef\Bs   = "051B  
\mathchardef\Bsi  = "0526  
\mathchardef\Bt   = "051C  
\mathchardef\Bu   = "051D  
\mathchardef\Bf   = "0527  
\mathchardef\Bff  = "051E  
\mathchardef\Bch  = "051F  
\mathchardef\Bps  = "0520  
\mathchardef\Bo   = "0521  
\mathchardef\Bome = "0524  
\mathchardef\BG   = "0500  
\mathchardef\BD   = "0501  
\mathchardef\BTh  = "0502  
\mathchardef\BL   = "0503  
\mathchardef\BX   = "0504  
\mathchardef\BP   = "0505  
\mathchardef\BS   = "0506  
\mathchardef\BU   = "0507  
\mathchardef\BF   = "0508  
\mathchardef\BPs  = "0509  
\mathchardef\BO   = "050A  
\mathchardef\BDpr = "0540  
\mathchardef\Bstl = "053F  

\newdimen\xshift \newdimen\xwidth \newdimen\yshift \newdimen\ywidth
\def\fline{\hbox to\hsize}
\def\ins#1#2#3{\vbox to0pt{\kern-#2pt\hbox{\kern#1pt #3}\vss}\nointerlineskip}

\def\eqfig#1#2#3#4#5{
\par\xwidth=#1pt \xshift=\hsize \advance\xshift
by-\xwidth \divide\xshift by 2
\yshift=#2pt \divide\yshift by 2
\fline{\hglue\xshift \vbox to #2pt{\vfil
#3 \includegraphics{#4.eps}
}\hfill\raise\yshift\hbox{#5}}}

\def\8{\write12}  

\def\figini#1{
\catcode`\%=12\catcode`\{=12\catcode`\}=12
\catcode`\<=1\catcode`\>=2
\openout12=#1.ps}

\def\figfin{
\closeout12
\catcode`\%=14\catcode`\{=1%
\catcode`\}=2\catcode`\<=12\catcode`\>=12}

\let\a=\alpha           \let\e=\varepsilon
      \let\th=\theta      \let\l=\lambda
    \let\n=\nu     \let\x=\xi     \let\p=\pi        \let\r=\varrho
 \let\t=\tau    \let\f=\varphi 
  \let\o=\omega

\let\dpr=\partial
\def\tende#1{\,\vtop{\ialign{##\crcr\rightarrowfill\crcr
 \noalign{\kern-1pt\nointerlineskip} \hskip3.pt${\scriptstyle
 #1}$\hskip3.pt\crcr}}\,}
\def\otto{\,{\kern-1.truept\leftarrow\kern-5.truept\to\kern-1.truept}\,}
\def\fra#1#2{{#1\over#2}}

\def\DD{{\cal D}}

\def\RR{{\cal R}}\def\TT{{\cal T}}

\def\ZZ{{\cal Z}}%
\def\V#1{{\bf#1}}%
\def\defi{\,{\buildrel def\over=}\,}%
\def\ig{\int}\def\io{\infty}%
\def\*{{\vskip2mm}}\def\0{\noindent}%
\def\iniz{\setcounter{equation}{0}}
\renewcommand{\theequation}{\arabic{section}.\arabic{equation}}%
\def\be{\begin{equation}}%
\def\ee{\end{equation}}%
\def\wt{\widetilde}

\def\fa{\forall\,}

\def\eg{{\it e.g.\ }\def\ie{{\it i.e.\ }}}

\begin{document}

\title{{\bf Perturbation theory}}

\author{Giovanni Gallavotti \\ {\it Dip. Fisica, Universit\`a di
Roma and  I.N.F.N., sezione Roma-1}}

\date{}
\maketitle

\section*{Article Outline}

\begin{enumerate}

\item Glossary

\item Definition 

\item Introduction

\item Poincar\'e's theorem and quanta

\item Mathematics and Physics. Renormalization

\item Need of convergence proofs

\item Multiscale analysis

\item A paradigmatic example of PT problem

\item Lindstedt series

\item Convergence. Scales. Multiscale analysis

\item Non convergent cases

\item Conclusion and Outlook

\item Bibliography

\end{enumerate}

\section{Definition of the subject and its importance}

{\bf Perturbation Theory:} Computation of a quantity depending on a
  parameter $\e$ starting from the knowledge of is value for $\e=0$ by
  deriving a power series expansion in $\e$, under the assumption of
  its existence, and if possible discussing the interpretation of the
  series. Perturbation theory is very often the only way to get a
  glimpse of the properties of systems whose equations cannot be
  ``explicitly solved'' in computable form. Its importance is
  witnessed by its applications in Astronomy, where it lead not only
  to the discovery of new planets (Neptune) but also to the disovery
  of Chaotic motions, with the completion of the Copernican revolution
  and the full understanding of the role of Aristotelian Physics
  formalized into uniform rotations of deferents and epicycles
  (nowadays Fourier representation of quasi periodic motions). It also
  played an essential role in the development of Quantum Mechanics and
  the understanding of the periodic table. The successes of Quantum
  Field Theory in Electrodynamics first, then in Strong interactions
  and finally in the unification of the elementary forces (stong,
  electromagnetic and weak) are also due to perturbation theory, which
  has also been essential in the theoretical understanding of the
  critical point universality. The latter two themes concern the new
  methods that have been developed in the last fifty years, marking a
  kind of new era for perturbation theory; namely dealing with
  singular problems, via the techniques called, in Physics,
  ``Renormalization Group'' and, in Mathematics, ``Multiscale
  Analysis''.

\section{Glossary}

\begin{itemize}

\item {\bf Formal power series}: a power series, giving the value
of a function $f(\e)$ of a parameter $\e$, that is derived
assuming that $f$ is analytic in $\e$.

\item {\bf Renormalization group}: method for multiscale analysis
and resummation of formal power series. Usually applied to define
a systematic collection of terms to organize a formal power series
into a convergent one.

\item {\bf Lindstedt Series}: an algorithm to develop formal power
series for computing the parametric equations of invariant tori in
systems close to integrable.

\item {\bf Multiscale Problem}: any problem in which an infinite
number of scales play a role.

\end{itemize}

 \*
\section{Introduction}\iniz

Perturbation theory, henceforth PT, arises when the value a function
of interest is associated with a problem depending on a parameter,
here called $\e$.  The value has to be a simple, or at least explicit
and rigorous, computation for $\e=0$ while its computation for
$\e\ne0$, small, is attempted by expressing it as the sum of a power
series in $\e$ which will be called here the ``solution''.

It is important to say since the beginning that a real PT solution of a
problem involves two distinct steps: the first is to show that
assuming that there is a convergent power series solving the problem
then the {\it coefficients} of the $n$-th power of $\e$ exist and can be
computed via finite computation. The resulting series will be called
{\it formal solution} or {\it formal series} for the problem. The
second step, that will be called {\it convergence theory}, is to prove
that the formal series converges for $\e$ small enough, or at
least find a ``summation rule'' that gives a meaning to the formal
series thus providing a real solution to the problem.  None of the two
problems is trivial, in the interesting cases, although the second is
certainly the key and a difficult one.

Once Newton's law of universal gravitation was established it became
necessary to develop methods to find its implications.  Laplace's
``M\'ecanique C\'eleste'', \cite{La799}, provided a detailed and
meticulous exposition of a general method that has become a classic,
if not the first, example of perturbation theory, quite different from
the parallel analysis of Gauss which can be more appropriately
considered a ``non perturbative'' development.

Since Laplace one can say that many applications along his lines
followed.  In the XIX century wide attention was dedicated to extend
Laplace's work to cover various astronomical problems: tables of the
coefficients were dressed and published, and algorithms for their
construction were devised, and planets were discovered (Neptune,
1846). Well known is the ``Lindstedt algorithm'' for the computation of
the $n$-th order coefficients of the PT series for the non resonant
quasi periodic motions. The algorithm provides a power series
representation for the quasi periodic motions with non resonant
frequencies which is extremely simple: however it represents the
$n$-th coefficient as a sum of many terms, some of size of the order
of a power on $n!$. Which of course is a serious problem for the
convergence.

It became a central issue, known as the ``small denominators
problem'' after Poincar\'e's deep critique of the PT method,
generated by his analysis of the three body problem. It led to his
``non-integrability theorem'' about the generic nonexistence of
convergent power series in the perturbation parameter $\e$ whose
sum would be a constant of motion for a Hamiltonian $H_\e$, member
of a family of Hamiltonians parameterized by $\e$ and reducing to
an integrable system for $\e=0$. The theorem suggested (to some)
that even the PT series of Lindstedt (to which Poincar\'e's
theorem does not apply) could be meaningless even though formally
well defined, \cite{Po892}.

A posteriori, it should be recognized that PT was involved also in the
early developments of Statistical Mechanics in the XIX century: the
virial theorem application to obtain the Van der Waals equation of
state can be considered a first order calculation in PT (although this
became clear only a century later with the identification of $\e$ as
the inverse of the space dimension).

\section{Poincar\'e's theorem and quanta}\iniz

With Poincar\'e begins a new phase: the question of convergence of
series in $\e$ becomes a central one in the Mathematics
literature. Much less, however, in the Physics literature where the
new discoveries in the atomic phenomena attracted the attention. It
seems that in the Physics research it was taken for granted that
convergence was not an issue: atomic spectra were studied via PT and
early authoritative warnings were simply disregarded (for instance,
explicit by Einstein, in \cite{Ei17}, and clear, in \cite{Fe923b}, but
``timid'' being too far against the mainstream, for his young age). In
this way quantum theory could grow from the original formulations of
Bohr, Sommerfeld, Eherenfest relying on PT to the final formulations
of Heisenberg and Schr\"odinger quite far from it.  Nevertheless the
triumph of quantum theory was quite substantially based on the
technical development and refinement of the methods of formal PT: the
calculation of the Compton scattering, the Lamb shift, Fermi's weak
interactions model and other spectacular successes came in spite of
the parallel recognition that the some of the series that were being
laboriously computed not only could not possibly be convergent but
their very existence, to all orders $n$, was in doubt.

The later Feynman graphs representation of PT was a great new tool
which superseded and improved earlier graphical representations of
the calculations. Its simplicity allowed a careful analysis and
understanding of cases in which even {\it formal} PT seemed
puzzlingly failing.

Renormalization theory was developed to show that the convergence
problems that seemed to plague even the computation of the individual
coefficients of the series, hence the formal PT series at fixed order,
were, in reality, often absent, in great generality, as suspected by the
earlier treatments of special (important) cases, like the higher order
evaluations of the Compton scattering, and other quantum
electrodynamics cross sections or anomalous characteristic constants
(\eg the magnetic moment of the muon).

\section{Mathematics and Physics. Renormalization}\iniz

In 1943 the first important result on the convergence of the series of
the Lindstedt kind was obtained by Siegel, \cite{Si43}: a formal PT
series, of interest in the theory of complex maps, was shown to be
convergent. Siegel's work was certainly a stimulus for the later work
of Kolmogorov who solved, \cite{Ko54}, a problem that had been
considered not soluble by many: to find the convergence conditions and
the convergence proof of the Lindstedt series for the quasi periodic
motions of a generic analytic Hamiltonian system, in spite of
Poincar\'e's theorem and actually avoiding contradiction with it. Thus
showing the soundness of the comments about the unsatisfactory aspects
of Poincar\'e's analysis that had been raised almost immediately by
Weierstrass, Hadamard and others.

In 1956 not only Kolmogorov theorem appeared but also convergence of
another well known and widely used formal series, the virial series,
was achieved in an unnoticed work by Morrey, \cite{Mo55}, and
independently rediscovered in the early 1960's.

At this time it seems that all series with well defined terms were
thought to be either convergent or at least asymptotic: for most
Physicists convergence or asymptoticity were considered of little
interest and matters to be left to Mathematicians.

However with the understanding of the formal aspects of
renormalization theory the interest in the convergence properties of
the formal PT series came back to the center of attention.

On the one hand mathematical proofs of the existence of the PT
series, for interesting quantum fields models, to all orders were
investigated settling the question once and for all (Hepp's
theorem, \cite{He66}); on the other hand it was obvious that even
if convergent (like in the virial or Meyer expansions, or in the
Kolmogorov theory) it was well understood that the radius of
convergence would not be large enough to cover all the physically
interesting cases. The sum of the series would in general become
singular in the sense of analytic functions and, even if admitting
analytic continuation beyond the radius of convergence, a
singularity in $\e$ would be eventually hit. The singularity was
supposed to correspond to very important phenomena, like the
critical point in statistical mechanics or the onset of chaotic
motions (already foreseen by Poincar\'e in connection with his non
convergence theorem).  Thus research developed in two direction.

The first aimed at understanding the nature of the singularities from
the formal series coefficients: in the 1960's many works achieved the
understanding of the scaling laws (\ie some properties of the
divergences appearing at the singularities of the PT series or of its
analytic continuation, for instance in the work of M.Fisher, Kadanoff,
Widom and may others). This led to trying to find {\it resummations},
\ie to collect terms of the formal series to transform them into {\it
convergent series} in terms of {\it new parameters}, the {\it
running couplings}.

The latter would be singular functions of the original $\e$ thus
possibly reducing the study of the singularity to the singularities of
the running couplings. The latter could be studied by independent
methods, typically by studying the iterations of an auxiliary
dynamical system (called the {\it beta function flow}). This was the
{\it approach} or {\it renormalization group method} of Wilson,
\cite{WK74,Ga01}.

The second direction was dedicated to finding out the real meaning of
the PT series in the cases in which convergence was doubtful or {\it a
priori} excluded: in fact already Landau had advanced the idea that
the series could be just illusions in important problems like the
paradigmatic quantum field theory of a scalar field or the fundamental
quantum electrodynamics, \cite{Fr82b,Gr99}.

In a rigorous treatment the function that the series were supposed to
represent would be in fact a trivial function with a dependence on
$\e$ unrelated to the coefficients of the well defined and non trivial
but formal series. It was therefore important to show that there were
at least cases in which the perturbation series of a nontrivial
problem had a meaning determined by its coefficients. This was studied
in the scalar model of quantum field theory and a proof of ``non
triviality'' was achieved after the ground breaking work of Nelson on
two dimensional models, \cite{Ne66,Si74}: soon followed by the similar
results in two dimensions and the difficult extension to three
dimensional models by Glimm and Jaffe, \cite{GJ81}, and generating
many works and results on the subject which took the name of
``constructive field theory'', \cite{Ga85b}.

But Landau's {\it triviality conjecture} was actually dealing with the
``real problem'', \ie the $4$-dimensional quantum fields.
The conjecture remains such at the moment, in spite of very intensive
work and attempts at its proof. The problem had relevance because it
could have meant that not only the simple scalar models of
constructive field theory were trivial but also the QED series which
had received strong experimental support with the correct prediction
of fine structure phenomena could be illusions, in spite of their well
defined PT series: which would be remain as mirages of a non
existing reality.

The work of Wilson made clear that the ``triviality conjecture'' of
Landau could be applied only to theories which, after the mentioned
resummations, would be controlled by a beta function flow that could
not be studied perturbatively, and introduced the new notion of {\it
asymptotic freedom}.  This is a property of the beta function flow,
implying that the running couplings are bounded and small so that
the resummed series are more likely to have a meaning, \cite{WK74}.

This work revived the interest in PT for quantum fields with attention
devoted to new models that had been believed to be non renormalizable.
Once more the apparently preliminary problem of developing a formal PT
series plaid a key role: and it was discovered that many Yang-Mills
quantum field theories were in fact renormalizable in the ultraviolet
region, \cite{HV72,Ho99}, and an exciting period followed with
attempts at using Wilson's methods to give a meaning to the Yang-Mills
theory with the hope of building a theory of the strong
interactions. Thus it was discovered that several Yang Mills theories
were asymptotically free as a consequence of the high symmetry of the
model, proving that what seemed to be strong evidence that no
renormalizable model would have asymptotic freedom was an ill founded
believe (that in a sense slowed down the process of understanding, and
not only of the strong interactions).

Suddenly understanding the strong interactions, until then considered
an impossible problem became possible, \cite{Gr99}, as solutions could
be written and {\it effectively computed} in terms of PT which,
although not proved to be convergent or asymptotic (still an open
problem in dimension $d=4$) were immune to the argument of Landau. The
impact of the new developments lead a little later to the unification
of all interactions into the {\it standard model} for the theory of
elementary particles (including the electromagnetic and weak
interactions). The standard model was shown be asymptotically free
{\it even in presence of symmetry breaking}, at least if a few other
interactions in the model (for instance the Higgs particle self
interaction) were treated heuristically while waiting for the
discovery of the ``Higgs particle'' and for a better understanding of the
structure of the elementary particles at length scales intermediate
between the Fermi scale ($\sim 10^{-15}\,cm$ (the weak interactions
scale) and the Planck scale (the gravitational interaction scale, $15$
orders of magnitude below).

Given that the very discovery of renormalizability of Yang-Mills
fields and the birth of a strong interactions theory had been
firmly grounded on experimental results, \cite{Gr99}, the latter
``missing step'' was, and still is, considered an acceptable gap.

\section{Need of convergence proofs}\iniz

The story of the standard model is paradigmatic of the power of PT: it
should convince anyone that the analysis of formal series, including
their representation by diagrams, which plays an essential part, is to
be taken seriously. PT is certainly responsible for the revival and
solution of problems considered by many as hopeless.

In a sense PT in the elementary particles domain can only, so far,
partially be considered a success. Different is the situation in the
developments that followed the works of Siegel and Kolmogorov. Their
relevance for Celestial Mechanics and for several problems in applied
physics (particle accelerator design, nuclear fusion machines for
instance) and for statistical mechanics made them too the object of a
large amount of research work.

The problems are simpler to formulate and often very well posed but
the possibility of existence of chaotic motions, always looming, made
it imperative not to be content with heuristic analysis and imposed
the quest of mathematically complete studies. The lead were the works
of Siegel and Kolmogorov. They had established convergence of certain
PT series, but there were other series which would certainly be not
convergent even though formally well defined and the question was,
therefore, which would be their meaning.

More precisely it was clear that the series could be used to find
approximate solutions to the equations, representing the motion for
very long times under the assumption of ``small enough'' $\e$. But
this could hardly be considered an understanding of the PT series in
Mechanics: the estimated values of $\e$ would have to be too small to
be of interest, with the exception of a few special cases. The real
question was what could be done to give the PT series the status of
exact solution.

As we shall see the problem is deeply connected with the above
mentioned asymptotic freedom: this is perhaps not surprising because
the link between the two is to be found in the ``multiscale
analysis'' problems, which in the last half century have been the core
of the studies in many areas of Analysis and in Physics,
when theoretical developments and experimental techniques became finer and
able to explore nature at smaller and smaller scales.

\section{Multiscale analysis}\iniz

To illustrate the multiscale analysis in PT it is convenient to
present it in the context of Hamiltonian mechanics, because in this
field it provides us with nontrivial cases of almost complete success.

We begin by contrasting the work of Siegel and that of Kolmogorov: which
are based on radically different methods. The first
being much closer in spirit to the developments of renormalization
theory and to the Feynman graphs.

Most interesting formal PT series have a common feature: namely
their $n$-th order coefficients are constructed as sums of many
``terms'' and the first attempt to a complete analysis is to
recognize that their sum, which gives the uniquely defined $n$-th
coefficient is much smaller than the sum of the absolute values of
the constituent terms. This is a property usually referred as a
``cancellation'' and, as a rule, it reflects some symmetry
property of the problem: hence one possible approach is to look
for expressions of the coefficients and for cancellations which
would reduce the estimate of the $n$-th order coefficients, very
often of the order of a power of $n!$, to an exponential estimate
$O(\r^{-n})$ for some $\r>0$ yielding convergence (parenthetically
in the mentioned case of Yang-Mills theories the reduction is even
more dramatic as it leads from divergent expressions to finite
ones, yet of order $n!$)

The multiscale aspect becomes clear also in Kolmogorov's method
because the implicit functions theorem has to be applied over and over
again and deals with functions implicitly defined on smaller and
smaller domains, \cite{Ga85,Ga86}. But the method purposedly avoids
facing the combinatorial aspects behind the cancellations so much
followed, and cherished, \cite{HV72}, in the Physics works.

Siegel's method was developed to study a problem in which no
grouping of terms was eventually needed, even though this was by
no means clear {\it a priori}, \cite{Po86}; and to realize that no
cancellations were needed forced to consider the problem as a
multiscale one because the absence of rapid growth of the $n$-th
order coefficients became manifest after a suitable ``hierarchical
ordering'' of the terms generating the coefficients. The approach
establishes a strong connection with the Physics literature
because the technique to study such cases was independently
developed in quantum field theory with renormalization, as shown
by Hepp in \cite{He66}, relying strongly on it.  This is very
natural and, in case of failure, it can be improved by looking for
``resummations'' turning the power series into a convergent series
in terms of functions of $\e$ which are singular but controllably
so. For details see below and \cite{Ga94b}.

What is ``natural'', however, is a very personal notion and it is not
surprising that what some consider natural is considered unnatural or
clumsy or difficult (or the three qualifications together) by others.

Conflict arises when the same problem can be solved by two different
``natural'' methods. and in the case of PT for Hamiltonian systems close
to integrable ones (closeness depending on the size of a parameter
$\e$), the so called ``small denominators'' problem, the methods of
Siegel and Kolmogorov are antithetic and an example of the just
mentioned dualism.

The first method, that will be called here ``Siegel's method'' (see
below for details), is based on a careful analysis of the structure of
the various terms that occur at a given PT order achieving a proof
that the $n$-th order coefficient which is represented as the sum of
many terms some of which {\it might} have size of order of a power of
$n!$ has in fact a size of $O(\r^{-n})$ so that the PT series is
convergent for $|\e|<\r$.  Although strictly speaking original work of
Siegel does not immediately apply to the Hamiltonian Mechanics
problems (see below), it can nevertheless be adapted and yields a
solution, as made manifest much later in \cite{Po86,El96,Ga94b}.

The second method, called here ``Kolmogorov's method'', instead
does not consider the individual coefficients of the various
orders but just regards the sum of the series as a solution of an
implicit function equation (a ``Hamilton-Jacobi'' equation) and
devises a recursive algorithm approximating the unknown sum of the
PT series by functions analytic in a disk of fixed radius $\r$ in
the complex $\e$-plane, \cite{Ga85,Ga86}.

Of course the latter approach implies that no matter how we achieve
the construction of the $n$-th order PT series coefficient there will
have to be enough cancellations, if at all needed, so that it turns out
bounded by $O(\r^{-n})$. And in the problem studied by Kolmogorov
cancellations would be necessarily present if the $n$-th order
coefficient was represented by the sum of the terms in the Lindstedt
series.

That this is not obvious is supported by the fact that it was considered
an open problem, for about thirty years, to find a way to exhibit
explicitly the cancellation mechanism in the Lindstedt series implied
by Kolmogorov's work. This was done by Eliasson, \cite{El96}, who
proved that the coefficients of the PT of a given order $n$ as
expressed by the construction known as the ``Lindstedt algorithm''
yielded coefficients of size of $O(\r^{-n})$: his argument, however,
did not identify in general which term of the Lindstedt sum for the
$n$-th order coefficient was compensated by which other term or
terms. It proved that the sum had to satisfy suitable relations, which
in turn implied a total size of $O(\r^{-n})$. And it took a few more
years for the complete identification, \cite{Ga94b}, of the rules to
follow in collecting the terms of the Lindstedt series which would
imply the needed cancellations.

It is interesting to remark that, aside from the example of
Hamiltonian PT, multiscale problems have dominated the development of
analysis and Physics in recent time: for instance they appear in
harmonic analysis (Carleson, Fefferman), in PDE's (DeGiorgi, Moser,
Caffarelli-Kohn-Ninberg) in relativistic quantum mechanics (Glimm,
Jaffe, Wilson) in Hamiltonian Mechanics (Siegel, Kolmogorov, Arnold,
Moser) in statistical mechanics and condensed matter (Fisher, Wilson,
Widom)... Sometimes, although not always, studied by PT techniques,
\cite{Ga01}.

\section{A paradigmatic example of PT problem}\iniz

It is useful to keep in mind an example illustrating technically what
it mean to perform a multiscale analysis in PT. And the case of quasi
periodic motions in Hamiltonian mechanics will be selected here, being
perhaps the simplest.

Consider the motion of $\ell$ unit masses on a unit circle and let
$\Ba=(\a_1,\ldots,\a_\ell)$ be their positions on the circle, \ie
$\Ba$ is a point on the on the torus $\TT^\ell=[0,2\p]^\ell$. The
points interact with a potential energy $\e f(\Ba)$ where $\e$ is a
strength parameter and $f$ is a trigonometric even polynomial, of
degree $N$: $f(\Ba)=\sum_{\Bn\in\ZZ^\ell,\,|\Bn|\le N} f_\Bn
e^{i\Bn\cdot\Ba},\, f_\Bn=f_{-\Bn}\in\RR$, where $\ZZ^\ell$ denotes
the lattice of the points with integer components in $\RR^\ell$ and
$|\Bn|=\sum_j|\n_j|$.

Let $t\to\Ba_0+\Bo_0 t$ be the motion with initial data, at time
$t=0$, $\Ba(0)=\Ba_0,\dot\Ba(0)=\Bo_0$, in which all particles rotate
at constant speed with rotation velocity
$\Bo_0=(\o_{01},\ldots,\o_{0\ell})\in\RR^\ell$. This is a solution for
the equations of motion for $\e=0$ and it is a quasi periodic
solution, \ie each of the angles $\a_j$ rotates periodically at
constant speed $\o_{0j}$, $j=1,\ldots,\ell$.

The motion will be called {\it non resonant} if the components of the
rotation speed $\Bo_0$ are rationally independent: this means that
$\Bo_0\cdot\Bn=0$ with $\Bn\in\ZZ^\ell$ is possible only if
$\Bn=\V0$. In this case the motion $t\to\Ba_0+\Bo_0 t$ covers, $\fa
\Ba_0$, densely the torus $\TT^\ell$ as $t$ varies.
The PT problem that we consider is to find whether there is a family
of motions ``of the same kind'' for each $\e$, small enough, solving
the equations of motion; more precisely whether there
exists a function $\V a_\e(\Bf),\,\Bf\in\TT^\ell$, such
that setting

\be\Ba(t)=\Bf+\Bo_0 t+\V a_\e(\Bf+\Bo_0 t),\qquad{\rm for}\
\Bf\in\TT^\ell\label{6.1}\ee
one obtains, $\fa \Bf\in \TT^\ell$ and for $\e$ small enough, a solution
of the equations of motion for a force $-\e\dpr_\Ba f(\Ba)$: \ie

\be \ddot
\Ba(t)=-\e \dpr_\Ba f(\Ba(t)).\label{6.2}\ee
By substitution of Eq. (\ref{6.1}) in Eq. (\ref{6.2}), the condition
becomes $(\Bo_0\cdot\dpr_\Bf)^2 \V a(\Bf+\Bo_0t)=-\dpr_\Ba
f(\Bf+\Bo_0t)$. Since $\Bo_0$ is assumed rationally independent
$\Bf+\Bo_0 t$ covers densely the torus $\TT^\ell$ as $t$ varies: hence
the equation for $\V a_\e$ is

\be(\Bo_0\cdot\dpr_\Bf)^2\V a_\e(\Bf)=\,-\e\,\dpr_\Ba f(\Bf+\V
a_\e(\Bf))\label{6.3}\ee
Applying PT to this equation means to look for a solution $\V
a_\e$ which is analytic in $\e$ small enough and in
$\Bf\in\TT^\ell$. In colorful language one says that the
perturbation effect is of slightly deforming a nonresonant torus
with given frequency spectrum ({\it i.e.} given $\Bo_0$) on which
the motion develops, without destroying it and keeping the quasi
periodic motion on it with the same frequency spectrum.

\section{Lindstedt series}\iniz

As it follows from a very simple special case of Poincar\'e's work,
Eq. (2) cannot be solved if also $\Bo_0$ is considered
variable and the dependence on $\e,\Bo_0$ analytic.  Nevertheless if
$\Bo_0$ is {\it fixed and non resonant} and if $\V a_\e$ is supposed
analytic in $\e$ small enough and in $\Bf\in\TT^\ell$, then there can
be at most one solution to the Eq. (2) with $\V a_\e(\V0)=\V0$ (which
is not a real restriction because if $\V a_\e(\Bf)$ is a solution also
$\V a_\e(\Bf+\V c_\e)+ \V c_\e$ is a solution for any constant $\V
c_\e$). This so because the coefficients of the power series in $\e$,
$\sum_{n=1}^\io \e^n \V a_n(\Bf)$, are uniquely determined if the
series is convergent. In fact they are trigonometric polynomials of
order $\le nN$ which will be written as

\be\Ba_n(\Bf)=\sum_{0<|\Bn|\le Nn}\Ba_{n,\Bn} e^{i\Bn\cdot\Bf}\label{7.1}\ee

It is convenient to express them in terms of graphs. The graphs to use
to express the value $\V a_{n,\Bn}$ are
\*

\0(i) trees with $n$ {\it nodes}
$v_1,\ldots,v_n$,
\\
(ii) one root $r$,
\\
(iii) $n$ lines joining pairs $(v'_i,v_i)$
of nodes or the root and one node, {\it always one and not more},
$(r,v_i)$;
\\
(iv) the lines will be different from each other and
distinguished by a mark label, $1,\ldots,n$ attached to them. The
connections between the nodes that the lines generate have to be {\it
loopless}, \ie the graph formed by the lines must be a tree.
\\
(v) The tree lines will be imagined oriented towards the root: hence a
partial order is generated on the tree and the line joining $v$ to
$v'$ will be denoted $\l_{v'v}$ and $v'$ will be the node closer to
the root immediately following $v$, hence such that $v'>v$ in the
partial order of tree.
\*

\eqfig{200}{130}{
\ins{-5}{71}{$r$}
\ins{-5}{91}{$\V e_r$}
\ins{14}{71}{$\Bn\kern-2pt=\kern-2pt\Bn_{\l_{0}}$}
\ins{24}{91}{$\l_{0}$}
\ins{50}{71}{$v_0$}
\ins{46}{91}{$\Bn_{v_0}$}
\ins{127}{100.}{$v_1$}
\ins{121.}{125.}{$\Bn_{v_1}$}
\ins{92.}{41.6}{$v_2$}
\ins{158.}{83.}{$v_3$}
\ins{191.7}{133.3}{$v_5$}
\ins{191.7}{100.}{$v_6$}
\ins{191.7}{71.}{$v_7$}
\ins{191.7}{-6.3}{$v_{11}$}
\ins{141.7}{20}{$v_{12}$}
\ins{191.7}{18.6}{$v_{10}$}
\ins{166.6}{42.}{$v_4$}
\ins{191.7}{54.2}{$v_8$}
\ins{191.7}{37.5}{$v_9$}
}{fig1}{}
\*

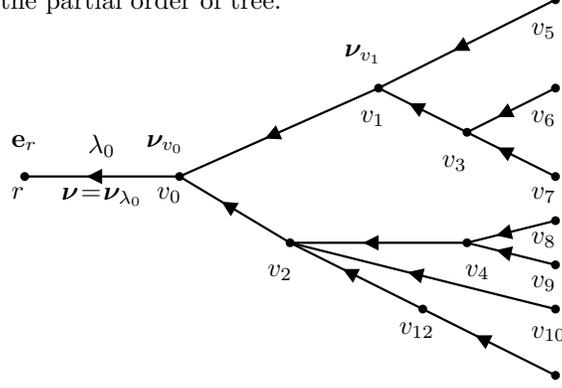
\0{\nota Fig.1: A tree $\th$ with
$m_{v_0}=2,m_{v_1}=2,m_{v_2}=3,m_{v_3}=2,m_{v_4}=2, m_{v_{12}}=1$
lines entering the nodes $v_i$, $k=13$. Some labels or decorations
explicitly marked (on the lines $\l_0,\l_1$ and on the nodes
$v_1,v_2$); the number labels, distinguishing the branches, are not
shown.  The arrows represent the partial ordering on the tree.\vfil}

The number of such trees is large and exactly equal to $n^{n-1}$,
as an application of Cayley's formula implies: their collection
will be denoted $T^0_n$.

To compute $\V a_{n,\Bn}$ consider all trees in $T^0_n$ and attach to
each node $v$ a vector $\Bn_v\in \ZZ^\ell$, called ``mode label'',
such that $f_{\Bn_v}\ne0$, hence $|\Bn_v|\le N$. To the root we
associate one of the coordinate unit vectors $\Bn_r\equiv\V e_r$. We
obtain a set $T_n$ of {\it decorated trees} (with $\le (2N+1)^{\ell
n}n^{n-1}$ elements, by the above counting analysis).

Given $\th\in T_n$ and $\l=\l(v',v)\in\th$ we define the {\it current}
on the line $\l$ to be the vector $\Bn(\l)\equiv
\Bn(v',v)\defi=\sum_{w\le v} \Bn_w$: \ie we imagine that the node
vectors $\Bn_{v_i}$ represent currents entering the node $v_i$ and
flowing towards the root. Then $\Bn(\l)$ is, for each $|l$, the sum of
the currents which entered all the nodes not following $v$, \ie
current accumulated after passing the node $v$.

The current flowing in the root line $\Bn=\sum_v\Bn_v$ will be denoted
$\Bn(\th)$.

Let $T^*_n$ be the set trees in $T_n$ in which {\it all lines} carry a
{\it non zero} current $\Bn(\l)\ne\V0$.  A {\it value} ${\rm
Val}(\th)$ will be defined, for $\th\in T^*_n$, by a product of node
factors and of line factors over all nodes and

\be{\rm Val}(\th)=\fra{i(-1)^n}{n!}\prod_{v\in\th} f_{\Bn_v} \prod_{\l=(v',v)}
\fra{\Bn_{v'}\cdot\Bn_v}{(\Bo_0\cdot\Bn(v',v))^2}\label{7.2}\ee
The coefficient $\V a_{n,\Bn}$ will then be

\be\V a_{n,\Bn}=\sum_{\th\in T^*_n\atop \Bn(\th)=\Bn} {\rm Val}(\th)
\label{7.3}\ee
and, when the coefficients are imagined to be constructed in this
way, the formal power series $\sum_{n=1}^\io \e^n \sum_{|\Bn|\le
Nn}\V a_{n,\Bn}$ is called the ``Lindstedt series''. Eq.
(\ref{7.2}) and its graphical interpretation in Fig.1 should be
considered the ``Feynman rules'' and the ``Feynman diagrams'' of
the PT for Eq. (\ref{6.3}), \cite{Ga95d,Ga01}.

\section{Convergence. Scales. Multiscale analysis.}\iniz

The Lindstedt series is well defined because of the non resonance
condition and the $n$-th term is not even a sum of too many terms:
if $F\defi\max_\Bn |f_\Bn|$,
each of them can be bounded by $\fra{F^n}{n!}  \prod_{\l\in\th}
\fra{N^2}{\Bo_0\cdot\Bn(\l)^2}$; hence their sum can be bounded, if $G$
is such that $\fra{(2N+1)^{\ell n}n^{n-1}F^n}{n!}\le G^n$, by $G^n
\prod_{\l\in\th} \fra{N^2}{\Bo_0\cdot\Bn(\l)^2}$.

Thus all $\V a_n$ are well defined and finite {\it but} the problem is
that $|\Bn(\l)|$ can be large (up to $Nn$ at given order $n$) and
therefore $\Bo_0\cdot\Bn(\l)$ although never zero can become very
small as $n$ grows.
For this reason the problem of convergence of the series is an example
of what is called a {\it small denominators problem}. And it is
necessary to assume more than just non resonance of $\Bo_0$ in order
to solve it in the present case: a simple condition is the {\it
Diophantine} condition, namely the existence of $C,\t>0$ such that

\be|\Bo_0\cdot\Bn|\ge
\fra{1}{C\,|\Bn|^\t},\qquad\fa\V0\ne\Bn\in\ZZ^\ell\label{8.1}\ee
But this condition is not sufficient in an obvious way: because it
only allows us to bound individual tree-values by $n!^a$ for some
$a>0$ related to $\t$; furthermore it is not difficult to check that
there are single graphs whose value is actually of ``factorial''
size in $n$. Although non trivial to see (as mentioned above) this was only
apparently so in the earlier case of Siegel's problem but it is the
new essential feature of the terms generating the $n$-th order
coefficient in Eq. (\ref{7.3}).

A resummation is necessary to show that the tree-values can be grouped
so that the sum of the values of each group can be bounded by
$\r^{-n}$ for some $\r>0$ and $\fa n$, although the group may contain
(several) terms of factorial size.  The terms to be grouped have to be
ordered hierarchically according to the sizes of the line factors
$\fra1{(\Bo_0\cdot\Bn(\l))^2}$, which are called {\it propagators} in
\cite{Ga94b,GBG04}.

A similar problem is met in quantum field theory where the graphs are
the {\it Feynman graphs}: such graphs can only have a small number of
lines that converge into a node but they can have loops, and to show
that the perturbation series is well defined to all orders it is also
necessary to collect terms hierarchically according to the propagators
sizes. The systematic way was developed by Hepp, \cite{He66,He69} for
the PT expansion of the Schwinger functions in quantum field theory of
scalar fields, \cite{Ga85b}. It has been used in many occasions later
and it plays a key role in the renormalization group methods in
Statistical Mechanics (for instance in theory of the ground state of
Fermi systems), \cite{BG95,Ga01}.

However it is in the Lindstedt series that the method is perhaps best
illustrated. Essentially because it ends up in a convergence proof,
while often in the field theory or statistical mechanics problems the PT
series can be only proved to be well defined to all orders, but they are
seldom, if ever, convergent so that one has to have recourse to other
supplementary analytic means to show that the PT series are asymptotic
(in the cases in which they are such).

The path of the proof is the following.
\*

\0(1) consider only trees in which no two lines $\l_+$ and $\l_-$,
    with $\l_+$ following $\l_-$ in the partial order of the tree,
    have the {\it same} current $\Bn_0$. In this case the maximum of
    the $\prod_\l\fra{1}{(\Bo_0\cdot\Bn(\l))^2}$ over all tress
    $\th\in T^*_n$ can be bounded by $G_1^n$ for some $G_1$.\\ This is
    an immediate consequence and the main result in the original
    Siegel's work, \cite{Si43}, which dealt with a different problem
    with small denominators in its formal PT solution: the
    coefficients of the series could also be represented by tree
    graphs, very similar too the ones above: but the only allowed
    $\Bn\in\ZZ^\ell$ were the non zero vectors with all components
    $\ge0$.  \\The latter property automatically guarantees that the
    graphs contain no pair of lines $\l_+,\l_-$ following each other
    as above in the tree partial order and having the same
    current. Siegel's proof also implies a multiscale analysis,
    \cite{Po86}: but it requires no grouping of the terms unlike the
    analogue Lindstedt series, Eq. (\ref{7.3}).  \*
\*

\0(2) Trees which contain lines $\l_+$ and $\l_-$, with $\l_+$
    following $\l_-$, in the partial order of the tree, and having the
    {\it same} current $\Bn_0$ can have values which have size of
    order $O(n!^a)$ with some $a>0$. Collecting terms is therefore
    essential.
\\ A
    line $\l$ of a tree is said to have scale $k$ if $2^{-k-1}\le
    \fra1{C|\Bo_0\cdot\Bn|}<2^{-k}$. The lines of a tree $\th\in
    T^*_n$ can then be collected in {\it clusters}.\footnote{\nota The
    scaling factor $2$ is arbitrary: any scale factor $>1$ could be used.}
\\ A cluster of scale $p$ is a maximal connected set of lines of scale
    $k\ge p$ with at least one line of scale $p$. Clusters are
    connected to the rest of the tree by lines of lower scale which
    can be {\it incoming} or {\it outgoing} with respect to the
    partial ordering. Clusters also contain nodes: a node is in a
    cluster if it is an extreme of a line contained in a cluster; such
    nodes are said {\it internal} to the cluster.

\figini{fig2}
\8</origine1assexper2pilacon|P_2-P_1| { 
\8<4 2 roll 2 copy translate exch 4 1 roll sub >
\8<3 1 roll exch sub 2 copy atan rotate 2 copy >
\8<exch 4 1 roll mul 3 1 roll mul add sqrt } def >
\8<>
\8</ellisse0 {
\8<exch dup 0 moveto 0 1 360 {>
\8<3 1 roll dup 3 2 roll dup 5 1 roll exch>
\8<5 2 roll  3 1 roll dup 4 1 roll sin mul>
\8<3 1 roll exch cos mul exch lineto} for stroke pop pop} def>
\8<>
\8</ellisse {
\8<gsave origine1assexper2pilacon|P_2-P_1| pop>
\8<ellisse0 grestore} def>
\8</punto { gsave >
\8<1.5 0 360 newpath arc fill stroke grestore} def >
\8<>
\8</puntone { gsave >
\8<4 0 360 newpath arc fill stroke grestore} def >
\8<>
\8</punta0 {0 0 moveto dup dup 0 exch 2 div lineto 0 >
\8<lineto 0 exch 2 div neg lineto 0 0 lineto fill >
\8<stroke } def >
\8<>
\8</dirpunta{
\8<gsave origine1assexper2pilacon|P_2-P_1| >
\8<0 translate 7 punta0 grestore} def >
\8<>
\8</linea {gsave 4 2 roll moveto lineto stroke grestore} def >
\8<>
\8</frecciac {0 index 3 index add 2 div 4 index 3 index >
\8<add 2 div exch 4 2 roll 5 index 5 index linea dirpunta} def >
\8<>
\8<gsave>
\8<1 0.8 scale>
\8</R {50} def>
\8</H {80} def>
\8</V {2 5 div R mul 3 5 div R mul} def>
\8</D {10 0} def>
\8<D translate>
\8<>
\8</P0 {R -1.5 div H} def>
\8</P1 {R 3 mul 6 div H} def>
\8</P2 {R 1.3 mul H} def>
\8</P3 {P1 V 3 -1 roll add 3 1 roll add exch} def>
\8</P4 {P2 V 3 -1 roll add 3 1 roll add exch} def>
\8</P5 {P4 2 R mul 5 div 0.2 R mul add 0 3 -1 roll add 3 1 roll>
\8< add 10 sub exch>
\8<} def>
\8</P6 {P4 2 R mul 5 div 2 R mul 2 div 3 -1 roll add 3 1 roll add -30 add>
\8<exch 15 sub  exch 2 5 div R mul add exch} def>
\8</P7 {3 R mul R 2 div add H 30 sub} def>
\8</PN {R 2.3 mul H 3 mul 4 div} def>
\8<R 2 div 0 translate>
\8<>
\8<P0 P2 linea>
\8<P1 P0 R 4 div 3 -1 roll add exch frecciac>
\8<P2 P1 frecciac>\8<P3 P1 frecciac>\8<P4 P2 frecciac>
\8<P5 P4 frecciac>\8<P6 P4 frecciac>
\8<
\8<PN P2 frecciac P7 PN frecciac>
\8<P0 puntone>\8<P1 punto>\8<P2 punto>\8<P3 punto PN punto>
\8<P4 punto >\8<P5 punto>\8<P6 punto>\8<P7 puntone>
\8<1.5 R mul 1.7 R mul P2 R 5 div add P2  R 5 div add >
\8<exch 10 add exch ellisse>
\8<R 2 div R 4 div P4 P4 exch 10 add exch ellisse>
\8<R 0.7 mul R 0.9 mul  P4 P4 exch 10 add exch  ellisse grestore>
\figfin
\*
\eqfig{160}{136}{
\ins{55}{60}{$v_{1}$}
\ins{110}{71}{$v_{2}$}
\ins{69}{93}{$v_{3}$}
\ins{134}{78}{$v_{5}$}
\ins{115}{117}{$v_{6}$}
\ins{120}{86}{$v_{4}$}
\ins{85}{33}{$T$}
\ins{130}{110}{$T'$}
\ins{100}{93}{$T''$}
\ins{148}{45}{$v_7$}
}{fig2}{}
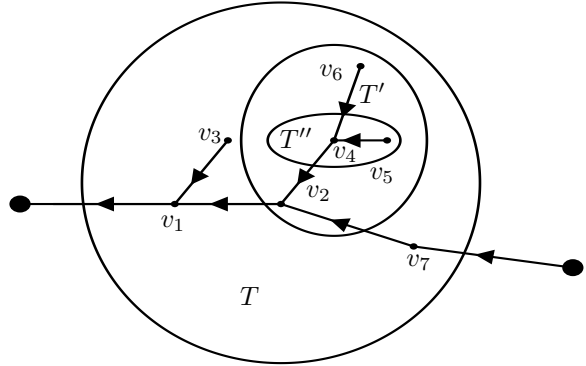
\0{\nota Fig.2: An example of three clusters symbolically delimited by
circles, as visual aids, inside a tree (whose remaining branches and
clusters are not drawn and are indicated by the bullets); not all
labels are explicitly shown.  The scales (not marked) of the branches
increase as one crosses inward the circles boundaries: recall,
however, that the scale labels are integers $\le1$ (hence typically
$\le0$).  The $\Bn$ labels are not drawn (but must be imagined).  If
the $\Bn$ labels of $(v_{4},v_{5})$ add up to $\V0$ the cluster $T''$
is a self-energy graph. If the $\Bn$ labels of
$(v_{2},v_{4},v_{5},v_{6})$ add up to $\V0$ the cluster $T'$ is a
self-energy graph and such is $T$ if the $\Bn$ labels of
$(v_1,v_2,v_3,v_{4},v_{5},v_{6},v_{7})$ add up to $\V0$. The cluster
$T'$ is maximal in $T$.\vfill}
\*

\0Of particular interest are the {\it self energy} clusters. These
    are clusters with only one incoming line and only one outgoing
    line which {\it furthermore} have the same current $\Bn_0$.
\0To simplify the analysis the Diophantine condition can be
    strengthened to insure that if in a tree graph the line incoming
    into a self energy cluster and ending in an internal node $v$ is
    detached from the node $v$ and reattached to another node internal
    to the same cluster which is not in a self-energy subcluster (if
    any) then the new tree nodes are still enclosed in the same
    clusters.  Alternatively the definition of scale of a line can be
    modified slightly to achieve the same goal.  \*

\0(3) Then it makes sense to sum together all the values of the trees
    whose nodes are collected into the same families of clusters and
    differ only because the lines entering the self energy clusters are
    attached to a different node internal to the cluster,
    but external to the inner self energy subclusters (if
    any). Furthermore the value of the trees obtained by changing
    simultaneously sign to the $\Bn_v$ of the nodes inside the self
    energy clusters have also to be added together.
\*

After collecting the terms in the described way it is possible to
check that each sum of terms so collected is bounded by $\r_0^{-n}$
for some $\r_0$ (which can also be estimated explicitly). Since the
number of addends left in not larger than the original one the bound
on $\sum_{\Bn}|\V a_{n,\Bn}|$ becomes $\le \fra{F^n (2N+1)^{\ell}n^n
N^{2n}}{n!}\r_0^{-n}\le \r^{-n}$, for suitable $\r_0,\r$, so that
convergence of the formal series for $\V a_\e(\Bf)$ is achieved for
$|\e|<\r$, see \cite{Ga94b}.

\section{Non convergent cases}\iniz

Convergence is {\it not} the rule: very interesting problems arise in
which the PT series is, or is believed to be, only asymptotic.  For
instance in quantum field theory the PT series are well defined but
they are not convergent: they can be proved, in the scalar $\f^4$ theories in
dimension $2$ and $3$ to be asymptotic series for a function of $\e$
which is {\it Borel summable}: this means in particular that the
solution can be in principle recovered, for $\e>0$ and small, just
from the coefficients of its formal expansion.

Other non convergent expansions occur in statistical mechanics, for
example in the theory of the ground state of a Fermi gas of particles
on a lattice of obstacles. This is still an open problem, and a
rather important one. Or occur in quantum field theory where sometimes they
can be proved to be Borel summable.

The simplest instances again arise in Mechanics in studying {\it resonant
  quasi periodic motions}. A paradigmatic case is provided by
  Eqs. (\ref{6.1}),(\ref{6.2}) when $\Bo_0$ has some vanishing
  components: $\Bo_0=(\o_1,\ldots,\o_r,0,\ldots,0)=(\wt\Bo_0,\V0)$
  with $1<r<\ell$. If one writes
  $\Ba=(\wt\Ba,\wt\Bb)\in\TT^r\times\TT^{\ell-r}$ and looks motions
  like Eq. (\ref{6.1}) of the form

\be \eqalign{
\wt\Ba(t)=&\wt\Bf +\wt\Bo_0 t+\wt {\V a}_\e(\wt\Bf +\wt\Bo_0 t)\cr
\wt\Bb(t)=&\Bb_0+\wt {\V b}_\e(\wt\Bf +\wt\Bo_0 t)\cr}\label{9.1}\ee
where $\wt {\V a}_\e(\wt\Bf),\wt {\V b}_\e(\wt\Bf)$ are functions of
$\wt\Bf\in \TT^r$, analytic in $\e$ and $\wt\Bf$.

In this case the analogue of the Lindstedt series can be devised
provided $\Bb_0$ is chosen to be a stationary point for the function
$\wt f(\wt\Bb)=\ig f(\wt\Ba,\wt \Bb)\fra{d\wt\Ba}{(2\p)^r}$, and
provided $\wt\Bo_0$ satisfies a Diophantine property
$|\wt\Bo_0\cdot\wt\Bn|>\fra1{C|\wt \Bn|^\t}$ for all $\V0\ne
\wt\Bn\in\ZZ^r$ and for $\t,C$ suitably chosen.

This time the series is likely to be, in general, non convergent
(although there is not a proof yet). And the terms of the Lindstedt
series can be suitably collected to improve the estimates. Nevertheless the
estimates cannot be improved enough to obtain convergence.
Deeper resummations are needed to show that in some cases the terms of
the series can be collected and rearranged into a convergent series.

The resummation is deeper in the sense that it is not enough to
collect terms contributing to a given order in $\e$ but it is
necessary to collect and sum terms of different order according to the
following scheme.

(1) the terms of the Lindstedt series are first ``regularized'' so
that the new series is manifestly analytic in $\e$ with, however, a
radius of convergence depending on the regularization. For instance
one can consider only terms with lines of scale $\le M$.

(2) terms of different orders in $\e$ are then summed together and
the series becomes a series in powers of functions $\l_j(\e;M)$ of
$\e$ with very small radius of convergence in $\e$, {\it but} with
an $M$-independent radius of convergence $\r$ in the $\l_j(\e,M)$.
The labels $j=0,1,\ldots,M$ are scale labels whose value is
determined by the order in which they are generated in the
hierarchical organization of the collection of the graphs
according to their scales.

(3) one shows that the functions $\l_j(\e;M)$ (``running couplings'')
can be analytically continued in $\e$ to an {\it $M$-independent
  domain $\DD$} containing the origin in its closure and where they
remain smaller than $\r$ for all $M$. Furthermore
$\l_j(\e;M)\tende{M\to\io}\l_j(\e)$, for $\e\in\DD$.

(4) the convergent power series in the running couplings admits an
asymptotic series in $\e$ at the origin which coincides with the
formal Lindstedt series.  Hence in the domain $\DD$ a meaning is
attributed to the sum of Lindstedt series.

(5) one checks that the functions $\wt{\V a}_\e,\wt{\V b}_\e$ thus
    defined are such that Eq. (\ref{9.1}) satisfies the equations of
    motion Eq. (\ref{6.1}).

The proof can be completed if the domain $\DD$ contains real points
$\e$.

If $\wt\Bb_0$ is a maximum point the domain $\DD$ contains a circle
tangent to the origin and centered on the positive real axis. So in
this case the $\wt{\V a}_\e,\wt{\V b}_\e$ are constructed in
$\DD\cap\RR_+$, $\RR_+\defi (0,+\io)$.

If instead $\wt\Bb_0$ is a minimum point the domain $\DD$ exists but
$\DD\cap\RR_+$ touches the positive real axis on a set of points with
positive measure and density $1$ at the origin. So $\wt{\V
a}_\e,\wt{\V b}_\e$ are constructed only for $\e$ in this set which
is a kind of ``Cantor set'',\cite{GG05}.

Again the multiscale analysis is necessary to identify the tree values
which have to be collected to define $\l_j(\e;M)$. In this case it is
an analysis which is much closer to the similar analysis that is
encountered in quantum field theory in the ``self energy resummations'',
which involve collecting and summing graph values of graphs contributing
to different orders of perturbation.

The above scheme can also be applied when $r=\ell$, \ie in the case of
the classical Lindstedt series when it is actually convergent: this
leads to an alternative proof of the Kolmogorov theorem which is
interesting as it is even closer to the renormalization group methods
because it expresses the solution in terms of a power series in running
couplings.\cite[Ch.8,9]{GBG04}.

\section{Conclusion and Outlook}\iniz

Perturbation theory provides a general approach to the solution of
problems ``close'' to well understood ones, ``closeness'' being
measured by the size of a parameter $\e$. It naturally consists of
two steps: the first is to find a formal solution, under the
assumption that the quantities of interest are analytic in $\e$ at
$\e=0$. If this results in a power series with well defined
coefficients then it becomes necessary to find whether the series
thus constructed, called {\it formal series}, converges.

In general the proof that the formal series exists (when it really
does) is nontrivial: typically in quantum mechanics problems (quantum
fields or statistical mechanics) this is an interesting and deep
problem giving rise to renormalization theory. Even in classical
mechanics PT of integrable systems it has been, historically, a
problem to obtain (in wide generality) the Lindstedt series (of which
a simple example is discussed above).

Once existence of a PT series is established, very often the series
is not convergent and at best is an asymptotic series. It becomes
challenging to find its meaning (if any, as there are
cases, even interesting ones, on which conjectures exist claiming that
the series have no meaning, like the quantum scalar field in dimension
$4$ with ``$\f^4$-interaction'' or quantum electrodynamics).

Convergence proofs, in most interesting cases, require a
multiscale analysis: because the difficulty arises as a consequence of
the behavior of singularities at infinitely many scales, as in the
case of the Lindstedt series above exemplified.

When convergence is not possible to prove, the multiscale analysis
often suggest ``resummations'', collecting the various terms whose sums yields
the formal PT series (usually the algorithms generating the PT series
give its terms at given order as sums of simple but many quantities, as
in the discussed case of the Lindstedt series). The collection involves
adding together terms of different order in $\e$ and results in a new
power series, the {\it resummed series}, in a family of parameters
$\l_j(\e)$ which are functions of $\e$, called the ``running
couplings'', depending on a ``scale index'' $j=0,1,\ldots$.

The running couplings are (in general) singular at $\e=0$ as functions
of $\e$ but $C^\io$ there, and obey equations that allow to study and
define them independently of a convergence proof. If the
running couplings can be shown to be so small, as $\e$ varies in a
suitable domain $\DD$ near $0$, to guarantee convergence of the
resummed series and therefore to give a meaning to the PT for
$\e\in\DD$ then the PT program can be completed.

The singularities in $\e$ at $\e=0$ are therefore all contained in
the running couplings, usually very few and the same for various
formal series of interest in a given problem.

The idea of expressing the sum of formal series as sum of convergent
series in new parameters, the running couplings, determined by other
means (a recursion relation denominated the beta function flow) is the
key idea of the renormalization group methods: PT in mechanics is a
typical and simple example.

On purpose attention has been devoted to PT in the analytic class: but
it is possible to use PT techniques in problems in which the functions
whose value is studied are not analytic; the techniques are somewhat
different and new ideas are needed which would lead quite far away
from the natural PT framework which is within the analytic class.

Of course there are many problems of PT in which the formal series are
simply convergent and the proof does not require any multiscale
analysis. However here attention having
been devoted to the novel aspect of PT that emerged in Physics and
Mathematics in the last half century and the
problems not requiring multiscale analysis have not been considered .
It is worth, however, to mention that even in simple convergent PT
cases it might be convenient to perform resummations. An example is
Kepler's equation

\be\ell=\x-\e\sin\x,\qquad \x,\ell\in T^1=[0,2\p]\label{10.1}\ee
which can be (easily) solved by PT. The resulting series has a
radius of convergence in $\e$ rather small (Laplace's limit):
however if a resummation of the series is performed transforming
it into a power series in a ``running coupling'' $\l_0(\e)$ (only
$1$, because no multiscale analysis is needed, the PT series being
convergent) given by, \cite[Vol. 2, p. 321]{LC}

\be \l_0\defi \fra{\e\,e^{\sqrt{1-\e^2}}}{1+\sqrt{1-\e^2}}.\label{10.2}\ee
The resummed series is a power series in $\l_0$ with radius of
convergence $1$ and when $\e$ varies between $0$ and $1$ the parameter
$\l_0$ corresponding to it goes from $0$ to $1$. Hence in terms of
$\l_0$ it is possible to invert {\it by power series} the Kepler
equation for all $\e\in[0,1)$, \ie in the entire interval of physical
interest (recall that $\e$ has the interpretation of eccentricity of
an elliptic orbit in the $2$-body problem). Resummations can improve
convergence properties.
\*

\section{Future directions}

It is always hard to indicate future directions, which usually turn
to different paths.  Perturbation theory is an ever evolving
subject: it is a continuous source of problems and its applications
generate new ones. Examples of outstanding problems are understanding
the triviality conjectures of models like quantum $\f^4$ field theory
in dimension $4$, \cite{Ga85b}; or a development of the theory of the
ground states of Fermionic systems in dimensions $2$ and $3$, \cite
{BG95}; a theory of weakly coupled Anosov flows to obtain information
of the kind that it is possible to obtain for weakly coupled Anosov
maps, \cite{GBG04}; uniqueness issues in cases in which PT series can
be given a meaning, but in a priori non unique way like the resonant
quasi periodic motions in nearly integrable Hamiltonian systems,
\cite{GBG04}.
\*

\0{\bf Acknowledgements:} Work partly supported by IHES.


\bibliographystyle{unsrt}

\end{document}